\documentclass{article}
\newcommand{\be}{\begin{equation}}
\newcommand{\ee}{\end{equation}}

\def\bea{\begin{eqnarray}}
\def\eea{\end{eqnarray}}
\def\bean{\begin{eqnarray*}}
\def\eean{\end{eqnarray*}}

\def\h{\huge}
\def\cl{\centerline}

\def\jrn#1#2#3#4{{#1} {\bf #2} (#4) #3}
\def\PRL{\it Phys. Rev. Lett.}
\def\PLB{\it Phys. Lett. B}
\def\PRD{\it Phys. Rev. D}
\def\NPB{\it Nucl. Phys. B}
\usepackage{amsmath}

\begin{document}
\begin{flushright}    UFIFT-HEP-04-09 \\ 
\end{flushright}
\vskip 2cm
\cl{\h {Chasing CHOOZ}\footnote{\uppercase{T}his work, based on an invited lecture at the 2003 Coral Gables Conference in December 2003,   was supported in part
by the US Department of Energy under grant DE-FG02-97ER41029}}
\vskip .5cm
\cl{\Large P. Ramond}
\vskip .5cm
\cl{Institute for Fundamental Theory, }
\cl{Physics Department, University of Florida} 
\cl{P.O. Box 118440, Gainesville, FL 32611, USA} 
\vskip 1cm 

\noindent We relate the MNS and CKM mixing matrices using ideas from grand unification. We catalog models in terms of the family symmetries of the down quark mass matrices, and emphasize the role of the Cabibbo angle in the lepton mixing matrix. We find a large class of models with an observable CHOOZ angle $~\sim \lambda/\sqrt{2}$.

\section{Introduction}
Since 1998 when SuperKamiokande announced incontrovertible  evidence of neutrino oscillations, a series of experiments have unravelled most of the fundamental neutrino parameters. Two important pieces of information still await determination: the absolute value of the masses, together with their hierarchy, and the third mixing angle, for which there only exists a limit set by the CHOOZ experiment. From the theoretical side, neutrino masses are hardly surprising in the context of the grand unified\cite{PS,SU5,SO10,E6} generalizations of the Standard Model. While some of us predicted one large mixing angle in the MNS lepton mixing matrix, the recent determination of {\it  two} large mixing angles comes as a surprise, and poses unforeseen theoretical challenges. In the following we offer several  remarks as to the theoretical implications of the present data set and note that with a modest infusion of grand unification ideas, it is possible to relate the CKM and MNS matrices. In a wide class of models, the CHOOZ angle is even predicted in terms of the Cabibbo angle.   

\section{The Data}
Five years of neutrino experiments\cite{SKatm,SKsol,SNO,CHOOZ,kamland,K2K}  can be succintly summarized\cite{FOGLI}: the MNS lepton mixing matrix 

$$\mathcal{ U}^{}_{MNS}~=~{\begin{pmatrix}\cos\phi_\odot&\sin\phi_\odot&\epsilon \\
-\cos\theta_\oplus~\sin\phi_\odot&\cos\theta_\oplus~\cos\phi_\odot&\sin\theta_\oplus \\
\sin\theta_\oplus~\sin\phi_\odot&-\sin\theta_\oplus~\cos\phi_\odot&\cos\theta_\oplus\end{pmatrix}}\ ,$$
has two large angles, one of which may be maximal. The size of the third  is limited by the CHOOZ\cite{CHOOZ} reactor experiment

$$ \sin^2 2\theta_\oplus~>~0.85\ ;\qquad  0.30~<~ \tan^2\phi_\odot~< ~0.65\ ;\qquad \vert~\epsilon~\vert^2_{}~<~0.005\ .$$
 It is  theoretically intriguing to produce a $(3\times 3)$ matrix with such  angles, and it  may be a powerful hint towards the resolution of the riddle of flavor. 
 Oscillation experiments determine only mass differences

$$\Delta m^2_\odot~=~\vert~m^2_{\nu_1}-m^2_{\nu_2}~\vert~\sim~7.\times 10^{-5}_{}~{\rm eV^2}$$
$$\Delta m^2_\oplus~=~\vert~m^2_{\nu_2}-m^2_{\nu_3}~\vert~\sim~3.\times 10^{-3}_{}~{\rm eV^2}\ ,$$
 but  WMAP\cite{WMAP} is the only experiment to set a limit on the absolute value of their masses

$$\sum_i~m^{}_{\nu_i}~<~.71~{\rm eV}$$
These allow for three three possible mass patterns\cite{SMIRNOV}:
\begin{itemize}
\item Hierarchical with $m^{}_{\nu_1}\prec m^{}_{\nu_2}\prec m^{}_{\nu_3}$
\item Inverted with $m^{}_{\nu_3}\prec m^{}_{\nu_2}\simeq m^{}_{\nu_1}$
\item Hyperfine with $m^{}_{\nu_1}\simeq m^{}_{\nu_2}\simeq m^{}_{\nu_3}$
\end{itemize}
In view of the strong theoretical reasons for linking quarks and leptons, the difference of their mixings is quite striking. In the context of the seesaw\cite{SEESAW} mechanism, these have possibly fundamental implications as to the structure of the electroweak singlet masses of the right-handed neutrinos.  In the following we  
explore the causes of these differences, and offer some theoretical prediction for the size of the CHOOZ angle. 

\section{CKM and MNS Mixings}
The $\Delta I_{\rm w}=1/2$ Higgs doublet breaking of electroweak symmetry yields the quark Yukawa Matrices

$$
\mathcal{ M}^{(2/3)}_{}~=\mathcal{ U}^{}_{2/3}\,
{\begin{pmatrix}m^{}_u&0&0\cr 0&m^{}_c&0\cr 0&0&m_t^{}\end{pmatrix}}
\,\mathcal{ V}^{\dagger}_{2/3}$$

$$
\mathcal{ M}^{(-1/3)}_{}~=~\mathcal{ U}^{}_{-1/3}\,
{\begin{pmatrix}m^{}_d&0&0\cr 0&m^{}_s&0\cr 0&0&m_b^{}\end{pmatrix}}
\,\mathcal{ V}^{\dagger}_{-1/3}
$$
from which the observable quark mixing matrix is deduced. 
 $$
\mathcal{ U}^{}_{CKM}~=~\mathcal{ U}^{\dagger}_{2/3}\,\mathcal{ U}^{}_{-1/3}
$$
Experimentally $\mathcal{ U}^{}_{CKM}$
 is nearly equal to the unit matrix, up to small powers of $\lambda$, the Cabibbo angle, allowing for    its Wolfenstein\cite{WOLF} expansion which uses the unit matrix as a starting point. The lesson here is that family mixing  is roughly the same for  charge $2/3$ and $-1/3$ quarks.  The charged leptons Yukawa Matrix

$$\mathcal{ M}^{(-1)}_{}~=~\mathcal{ U}^{}_{-1}\,
{\begin{pmatrix} m^{}_e&0&0\cr 0&m^{}_\mu&0\cr 0&0&m_\tau^{}\end{pmatrix}}
\,\mathcal{ V}^{\dagger}_{-1}$$
has hierarchical mass eigenvalues, as in the quark sector. To explain neutrino masses, it is simplest  to add one  right-handed neutrino per family. This implies a neutral leptons Yukawa Matrix

 {$$\mathcal{ M}^{(0)}_{ {Dirac}}~=~\mathcal{ U}^{}_{0}\,\mathcal{ D}_0^{}\,
\mathcal{ V}^{\dagger}_{0}~=~\mathcal{ U}^{}_{0}\,
{\begin{pmatrix} m^{}_1&0&0\cr 0&m^{}_2&0\cr 0&0&m_3^{}\end{pmatrix} }\,\mathcal{ V}^{\dagger}_{0}
$$}
and a lepton mixing matrix that is similar to the CKM matrix
$$\mathcal{ U}^{}_{MNS}~=~ \mathcal{ U}^{\dagger}_{-1}\,\mathcal{ U}^{}_{0}\,?$$
Can such a matrix be so different from its sister matrix in the quark sector? 
Most right-thinking theorists think that the right-handed neutrino masses 

$$\mathcal{ M}^{(0)}_{ {Majorana}}~\sim~ \Delta I_{\rm w}^{}=0$$
radically change the landscape: unlike  those of quarks and charged leptons,  the right-handed neutrinos  masses are unconstrained by electroweak quantum numbers. They only have lepton number, and since there are very good limits on lepton number conservation, one expects these masses to be very large. This leads to the seesaw\cite{SEESAW} mechanism, according to which the neutrino mass matrix is

$$\mathcal{ M}^{(0)}_{ {Seesaw}}~=~\mathcal{ M}^{(0)}_{ {Dirac}}\,
\frac{1}{\mathcal{ M}^{(0)}_{ {Majorana}}}\,\mathcal{ M}^{(0)\,t}_{ {Dirac}}\ .$$
We can rewrite it as

$$
\mathcal{ M}^{(0)}_{ {Seesaw}}~=~\mathcal{ U}^{}_{0}\,
 {\mathcal{ C}}\,\mathcal{ U}^{t}_{0}\ ,$$
where we have introduce the Central Matrix

$$ \mathcal{ C}~\equiv~\mathcal{ D}_0^{}\,\mathcal{ V}^{\dagger}_{0}\,
\frac{1}{\mathcal{ M}^{(0)}_{ {Majorana}}}\,\mathcal{ V}^{*}_{0}\,\mathcal{ D}_0^{}\ .$$
 It is diagonalized by a unitary matrix $\mathcal F$

$$ \mathcal{ C}~=~ \mathcal{ F}\,\mathcal{ D}^{}_\nu\, \mathcal{ F}^{\,t}_{}\ ,$$
while the physical neutrino masses are contained in the diagonal $
\mathcal{ D}_\nu^{}$. 
This alters the observable MNS lepton mixing matrix to

$$\mathcal{ U}^{}_{MNS}~=~ \mathcal{ U}^{\dagger}_{-1}\,\mathcal{ U}^{}_{0}\, \mathcal{ F}\ .$$
The seesaw adds to  the MNS matrix   the extra unitary matrix $ \mathcal{ F}$. 
This suggests we put models of neutrino masses in three  classes: 
 
\begin{itemize}
\item Models of type 0 for which   $ \mathcal{ F}$ contains no large angle,
\item Models of type I for which  only one large angle is in $ \mathcal{ F}$, 
\item Models of type II, for which both large angles reside in $ \mathcal{ F}$.
\end{itemize}
Type I models seem more generic to us than type II, since it is natural to have matrices with one, three or no large mixing angles.  

\section{A Pinch of Grand Unification}
If we want to relate the quark and lepton observables, we need to add some theoretical prejudice. Such is not hard to find, as it is made obvious both by anomaly cancellation and the quantum number structures: grand unification. While there is a great deal of uncertainty as to the nature of the grand unified group and the mechanism by which its large symmetries are broken, there is no doubt that  such a group is to be found (cosmologically speaking) in our distant past. It is therefore important to study the patterns implied.

Grand Unification relates $\Delta I^{}_{\rm w}=1/2$ quark and lepton Yukawa matrices. In the simplest case, $ SU(5)$,  we have

$$\mathcal{ M}^{(-1/3)}_{}~\sim~\mathcal{ M}^{(-1)\,t}_{}\ ,$$
where $t$ is the transpose, while in $ SO(10)$ which naturally includes right-handed neutrinos, 

$$\mathcal{ M}^{(2/3)}_{}~\sim~\mathcal{ M}^{(0)}_{ {Dirac}}\ .$$
Assuming only these two simple patterns,  we infer

$$\mathcal{ U}^{}_{-1/3}~\sim~\mathcal{ V}^{*}_{-1}\ ,\qquad \mathcal{ U}^{}_{2/3}~\sim~\mathcal{ U}^{}_{0}\ .$$
Putting these back in the MNS matrix, we find

\bea\nonumber\mathcal{ U}^{}_{MNS}&=& \mathcal{ U}^{\dagger}_{-1}\,\mathcal{ U}^{}_{-1/3}\,\mathcal{ U}^{\dagger}_{CKM}\, \mathcal{ F}\\
\nonumber&=& \mathcal{ V}^t_{-1/3}\,\mathcal{ U}^{}_{-1/3}\,\mathcal{ U}^{\dagger}_{CKM}\, \mathcal{ F}\ ,\eea
suggesting how the CKM matrix and its  Cabibbo expansion  enter into the lepton mixing observables.  It is interesting to note that it is the  structure 
of the charge $-1/3$ matrix that plays a dominant role in relating quark to lepton mixings.

\begin{itemize}
\item If  $\mathcal{ M}^{}_{-1/3}$ is family-symmetric,  

$$\mathcal{ U}^{}_{-1/3}~=~\mathcal{ V}^*_{-1/3}\ ,$$
and we get

$$\mathcal{ U}^{}_{MNS}~=~ \mathcal{ U}^{\dagger}_{CKM}\,\, \mathcal{ F}\ .$$
Both large angles must be in $\mathcal{ F}$; these are type II models. 

\item 
 If  $\mathcal{ M}^{}_{-1/3}$ is not family-symmetric,  one or two large angle can come from it, leading to type O and type I models.

\end{itemize}
Type O models require the two mixing angles among the right-handed down quarks to be uniformly large. Although logically possible, I will not discuss these models in the following.

There is a framework in which type I Models appear naturally. Start from  the Yukawa matrices, assuming that the masses a Cabibbo expansion for the mass ratios, as well as the Froggatt-Nielsen\cite{FN} scheme for the exponents, 

$$\mathcal{ M}^{(-1/3)}_{}~=~{\begin{pmatrix}  {\lambda^4_{}}& {\lambda^3_{}}& {\lambda^3_{}}\cr
 {\lambda^3_{}}& {\lambda^2_{}}& {\lambda^2_{}}\cr
 {\lambda}&a&b\end{pmatrix}}\ ,$$
and

$$\mathcal{ U}^{}_{CKM}~=~{\begin{pmatrix} 1& {\lambda}& {\lambda^3_{}}\cr  {\lambda}&1& {\lambda^2_{}}\cr  {\lambda^3_{}}& {\lambda^2_{}}&1\end{pmatrix}}\qquad \frac{m^{}_s}{m^{}_b}~\sim~ {\lambda^2_{}}\qquad \frac{m^{}_d}{m^{}_b}~\sim~ {\lambda^4_{}}\ .$$
It is simplest to consider  the no-mixing limit $(\lambda~\rightarrow~ 0)$ where

$$\mathcal{ M}^{(2/3)}_{}~=~{\begin{pmatrix} 0&0&0\cr 0&0&0\cr 0&0&m_t^{}\end{pmatrix} }\ ,\qquad 
\mathcal{ M}^{(-1/3)}_{}~=~{\begin{pmatrix} 0&0&0\cr 0&0&0\cr 0&a&b\end{pmatrix} }$$
 and 

 $$\mathcal{ U}^{}_{CKM}~=~\mathcal{ U}^{\dagger}_{2/3}\,\mathcal{ U}^{}_{-1/3}~=~1$$
Then $SU(5)$ suggest a non-symmetric charge $-1$ matrix as well, 

$$\mathcal{ M}^{(-1)}_{}~=~{\begin{pmatrix} 0&0&0\cr 0&0&a\cr 0&0&b\end{pmatrix}}\ ,$$
 so that 

$$\mathcal{ U}^{}_{-1}~=~{\begin{pmatrix} 1&0&0\cr 0&\cos\theta&\sin\theta\cr 0&-
\sin\theta&\cos\theta\end{pmatrix}}\ ,\qquad \tan\theta~=~\frac{a}{b}\ ,$$
contains one unsuppressed mixing angle. 

 $$\mathcal{ U}^{}_{MNS}~=~
{\begin{pmatrix} 1&0&0\cr 0&\cos\theta&\sin\theta\cr 0&-\sin\theta&\cos\theta\end{pmatrix}}\, \mathcal{ F}\ .$$
It is  natural from this point of view\cite{ILR,KING} to expect unsuppressed atmospheric neutrino mixing, although this approach does not fix the value of the angle.  

The relation between the CKM and MNS matrices suggest that the Cabibbo angle plays an important role and should not be neglected. In particular it could very well be that the CHOOZ angle is {\it solely} a Cabibbo effect. In type I Models, this hypothesis suggests that

$$\mathcal{ U}^{}_{MNS}~=~{\begin{pmatrix} 1& {\lambda^\alpha}& {\lambda^\beta}\cr
 {\lambda^\alpha}&
\cos\theta&\sin\theta\cr  {\lambda^\beta}& -\sin\theta&\cos\theta\end{pmatrix}}
~{\begin{pmatrix} \cos\phi&\sin\phi& {\lambda^\gamma}\cr -\sin\phi&\cos\phi& {\lambda^\delta}\cr  {\lambda^\gamma}& {\lambda^\delta}&1\end{pmatrix}}\ ,$$
where the exponents are completely unknown. As a result the CHOOZ angle is given by $ \lambda$ to an unknown power with an unknown prefactor, and our hypothesis is not very useful.

However, in type II models, where the $\Delta I_w=1/2$ Yukawa matrices are family symmetric, we find a more definitive prediction

$$\mathcal{ U}^{}_{MNS}={\begin{pmatrix} 1& {\lambda}& {\lambda^3}\cr
 {\lambda}&
1& {\lambda^2}\cr  {\lambda^3}&  {\lambda^2}&1\end{pmatrix}}
{\begin{pmatrix} \cos\phi&\sin\phi& {\lambda^\gamma}\cr
-\cos\theta~\sin\phi&\cos\theta~\cos\phi&\sin\theta\cr
\sin\theta~\sin\phi&-\sin\theta~\cos\phi&\cos\theta\end{pmatrix}}\ ,
$$
so that plausibly 

$${\rm CHOOZ~~ angle}~~\sim ~~ {\lambda}~ {\sin\theta}~\sim~\frac{ {\lambda}}{\sqrt{2}}\ ,$$
assuming  $\gamma<2$. This conclusion is particularly exciting as it can be tested in the foreseeable future; if correct it will also open the way to observable CP violation in neutrino physics!

We end this section with two additional remarks: one is that it will be desirable to fix a Wolfenstein parametrization for the MNS matrix. The second is that there is another class of models where the charge $-1/3$ Yukawa matrix is not family-symmetric. These are Family Cloning\cite{BR} models where  each family starts with its own gauge group. A tri-chiral order parameter naturally explains 
$$SU(3)_1\times SU(3)_2\times SU(3)_3~\rightarrow~SU(3)_{1+2+3}\ ,$$
but falls short in unifying the thre weak isospins into one

$$SU(2)_1\times SU(2)_2\times SU(2)_3~\rightarrow~SU(2)_{1+2}\times SU(2)_3\ ,$$
resulting in a Standard Model with lopsided gauge symmetry

$$SU(2)_{1+2}\times SU(2)_3\times SU(3)_{1+2+3}\ ,$$
yielding asymmetric Yukawa matrices and one large angle in $U_{-1}$.
\section{Large Angles and $\mathcal F$}
Finally we would like to discuss the theoretical implications of large angles in $\mathcal F$.  A$(3\times 3)$ matrix generically contains one or three large angles, but not two. From this point of view type I models are more desirable. 

As in our earlier work\cite{DLR}, consider a simplified case with  two families. Recall that 

$$ \mathcal{ C}~\equiv~ {\mathcal{ D}_0^{}}\,\mathcal{ V}^{\dagger}_{0}\,
\frac{1}{\mathcal{ M}^{(0)}_{ {Majorana}}}\,\mathcal{ V}^{*}_{0}\, {\mathcal{ D}_0^{}}\ ,$$
and that the $\Delta I_w=1/2$ Neutral Dirac Mass is hierarchical:
 
$$
\mathcal{ D}^{}_0~=~m{\begin{pmatrix} a\, {\lambda^\alpha_{}}&0\cr 0&1\end{pmatrix}}\ .$$
 If we call $M_1$ and $M_2$ the right-handed Majorana masses, simple algebra yields the central matrix

$$
 \mathcal{ C}~=~{\begin{pmatrix} (\frac{c^2}{M_1}+\frac{s^2}{M_2})\,a^2_{}\,
 {\lambda^{2\alpha}_{}}&(\frac{c\,s}{M_1}-\frac{c\,s}{M_2})\,a\, {\lambda^\alpha_{}}\cr
(\frac{c\,s}{M_1}-\frac{c\,s}{M_2})\,a\, {\lambda^\alpha_{}}&
(\frac{s^2}{M_1}+\frac{c^2}{M_2})\end{pmatrix}}\ ,
$$
where $s\ , c$ are the sine and cosine of a mixing angle.  

It is diagonalized by a  large mixing angle in one of two cases:

\begin{itemize}
\item $
 {\mathcal{ C}_{11}} ~\sim~  {\mathcal{ C}_{22}} ~\sim~  {\mathcal{ C}_{12}}\ . $
 This implies  

$$ s~\sim~b\, {\lambda^\alpha_{}}\ ,\qquad c~\sim~1\ ,$$
and 

$$
\frac{M_1}{M_2}~\sim~ {\lambda^{2\beta}_{}}\ .$$
Hence the Majorana masses must be hierarchical: there is a correlated hierarchy between the $\Delta I_w=1/2$ and $\Delta I_w=0$ sectors. Then

$$ \mathcal{ C}~=~ {\lambda^{2\alpha}_{}}\,\frac{m^2}{M^{}_1}\,
{\begin{pmatrix} {a^2}&{a\,b}\cr {a\,b}&{b^2}\end{pmatrix}}\ ,~~~\beta>\alpha$$

\item $
\mathcal{ C}_{11}\, , \,\mathcal{ C}_{22} ~\prec~ \mathcal{ C}_{12}
$. This is the level crossing case. We obtain 

$$
\lambda^\alpha ~\prec~ s ~\preceq~ \lambda^{\alpha-\beta}
$$
so that 
$$\mathcal{ C}~=~\frac{\lambda^\alpha~m^2}{\sqrt{-M_1 M_2}}\,
{\begin{pmatrix} 0&a\cr a&0\end{pmatrix}}\ ,$$
naturally leading to maximal mixing. Could the right-handed neutrinos of two families be Dirac partners?

\end{itemize}

\section{Conclusions}
The neutrino data set presents new theoretical challenges and hopefully hints towards the resolution of the family puzzle.  It points  to family mixing at very high scales and a  hierarchy among the right-handed neutrinos, in accordance with the grand-unified paradigm. This is a welcome feature in terms of modern theories of leptogenesis. 

We close with some suggestions to model builders. Go directly to the Planck scale (or just below). There you will find branes awaiting you with Weyl fermions we can right-handed neutrinos. Their interactions and masses are key to understanding the family riddles and symmetries, and even better laboratory neutrino physics opens  a window to their masses. Only then should you worry about the Standard Model; and for that you can always use $SO(10)$ or any such model\cite{AB,SOTEN}. 

\section{Acknowledgements} This is the first Coral Gables Conference without Behram Kursunoglu, and many of us who have attended countless times miss him. I shall always remember him as a charming advocate and tireless promoter of fundamental science.  Thanks to his efforts,  physicists from all over the world came yearly  to South Florida to discuss physics in a relaxed and gracious atmosphere.  I also wish to thank Drs A. Datta and  L. Everett   for useful discussions.

\end{document}